\documentclass[sigconf]{acmart}

\usepackage{booktabs, multirow} 
\usepackage[T1]{fontenc}
\usepackage{textcomp}
\usepackage{amsmath}
\usepackage{array}
\usepackage{breqn}
\usepackage{graphicx}
\usepackage{subcaption}
\usepackage{bm}
\usepackage{float}
\usepackage{amssymb, amsthm, enumitem}
\usepackage{hyperref}
\graphicspath{{figures/}}

\newcounter{para}

\newcommand{\hide}[1]{}

\newcolumntype{C}[1]{>{\centering\let\newline\\\arraybackslash\hspace{0pt}}m{#1}}

\newcommand\newdagger{\mathrel{\ooalign{$\dagger$\cr\hidewidth$\ddagger$\hidewidth\cr}}}

\newcommand{\extra}{\textsc{Extra}}
\newcommand{\hh}[1]{{\small\color{red}{\bf hh: #1}}}
\newcommand{\lli}[1]{{\small\color{blue}{\bf LL: #1}}}
\newcommand{\qh}[1]{{\small\color{brown}{\bf QH: #1}}}

\begin{document}
	\setlength{\textfloatsep}{2pt}
	\setlength{\abovedisplayskip}{2pt}
	\setlength{\belowdisplayskip}{2pt}
	\setlength{\parindent}{1em}
	\title{\extra: Explaining Team Recommendation in Networks}
	\author{Qinghai Zhou$^\dagger$,~~~Liangyue Li$^\dagger$,~~~Nan Cao$^\ddagger$,~~~Norbou Buchler$^\newdagger{}$ and Hanghang Tong$^\dagger$}
	\affiliation{\vspace{-4mm}
		\institution{$^\dagger$Arizona State University, $^\ddagger$Tongji University, $^\newdagger{}$ARL\\
			\vspace{-4mm}
			$^\dagger$\{qinghai.zhou, liangyue, hanghang.tong\}@asu.edu, $^\ddagger$nan.cao@gmail.com, $^\newdagger{}$norbou.buchler.civ@mail.mil}}
	\hide{\author{Qinghai Zhou$^\dagger$,~~~Liangyue Li$^\dagger$,~~~Nan Cao$^\ddagger$,~~~Norbou Buchler$^\newdagger{}$ and Hanghang Tong$^\dagger$}
	    \affiliation{
	        \institution{$^\dagger$Arizona State University, $^\ddagger$Tongji University, $^\newdagger{}$ARL\\\vspace{-4mm}
	            	     $^\dagger$\{qinghai.zhou, liangyue, hanghang.tong\}@asu.edu, $^\ddagger$nan.cao@gmail.com, $^\newdagger{}$norbou.buchler.civ@mail.mil}}}
	%
	\renewcommand{\shortauthors}{Q. Zhou et al.}
	\begin{abstract}


\hide{\hh{QH: check if SIGIR demo is double-blind?}}

State-of-the-art in network science of teams offers effective recommendation methods to answer questions like {\em who} is the best replacement, {\em what} is the best team expansion strategy, but lacks intuitive ways to explain {\em why} the optimization algorithm gives the specific recommendation for a given team optimization scenario. To tackle this problem, we develop an interactive prototype system,~\extra, as the first step towards addressing such a sense-making challenge, through the lens of the underlying network where teams embed, to explain the team recommendation results. The main advantages are (1) {\it Algorithm efficacy:} we propose an effective and fast algorithm to explain random walk graph kernel, the central technique for networked team recommendation; (2) {\it Intuitive visual explanation:} we present intuitive visual analysis of the recommendation results, which can help users better understand the rationality of the underlying team recommendation algorithm.

	\end{abstract}
	
	\hide{
		
	}
	\vspace{-6mm}
	\copyrightyear{2018} 
	\acmYear{2018} 
	\acmConference[RecSys '18]{Twelfth ACM Conference on Recommender Systems}{October 2--7, 2018}{Vancouver, BC, Canada}
	\acmBooktitle{Twelfth ACM Conference on Recommender Systems (RecSys '18), October 2--7, 2018, Vancouver, BC, Canada}
	\acmDOI{10.1145/3240323.3241610}
	\acmISBN{978-1-4503-5901-6/18/10}
	\maketitle
	
	\vspace{-2mm}
\section{Introduction}

The emergence of network science has been significantly changing the landscape of team-based research in recent years. 
A cornerstone behind team recommendation algorithms for various scenarios (e.g., {\em team member replacement}, {\em team expansion} and {\em team shrinkage}) is random walk graph kernel~\cite{borgwardt2007fast}. For instance, in team member replacement,  by applying random walk graph kernel to the team networks before and after replacement, it encodes both the skill match, structure match as well as the interaction between the two during the replacement process~\cite{li2015replacing}. Team member replacement further enables other team recommendation scenarios~\cite{li2017enhancing}. \hide{Although being effective in answering questions like {\em who} is the best replacement, {\em what} is the best team expansion strategy,} However, these existing methods lack intuitive ways to explain {\em why} the underlying algorithm gives the specific recommendation for a given team optimization scenario.

\noindent\textbf{Purpose and Novelty:} In this paper, we present an interactive prototype system,~\extra, a paradigm shift from {\em what} to {\em why}, for the purpose of explaining the networked team recommendation results. The novelties of \extra{} are (1) on the algorithmic side, we propose an effective and efficient algorithm to explain random walk graph kernel, where the key idea is to identify the most influential network substructures and/or attributes whose removal or perturbation will impact the graph kernel/similarity the most, and (2) on the system side, \extra{} is able to provide informative visual analysis for explaining various team recommendation scenarios from a variety of perspectives (e.g., edges, nodes and attributes). For example, given a candidate for team member replacement, we are able to tell (a) what the key connections are between the candidate and the existing team members, and (b) what the key skills the candidate has that might make him/her a potentially good replacement.

	\vspace{-2mm}
\section{Overview of  the System} \label{section2}

\begin{table*}[ht]\label{tab:summary}
	\centering
	\small
	\captionsetup{justification=centering}
	\begin{tabular}{|C{1.5cm}|C{4.9cm}|C{5.3cm}|C{4.7cm}|}
		\hline
		& {\bf Team Replacement} & {\bf Team Expansion} & {\bf Team Shrinkage} \\
		\hline
		{\bf Edges} & important common collaborations shared by the candidate and the departure member & new collaborations that the new member might establish & the most important but lacking collaborations the candidate should have\\
		\hline
		{\bf Nodes} & key existing team members both candidate and departure member collaborate with & key existing team members the new member will work with & key existing team members that the candidate should have collaborated with\hide{\qh{consider overall collaboration is not good}}\\
		\hline
		{\bf Attributes} & common and important skills shared by the candidate and the departure member  &  the unique skills the new team member brings that are critical to the team's new need & the most important skills that the candidate lacks \\
		\hline
	\end{tabular}
	\centering
	\vspace{1mm}
	\caption{Summary of system functionalities. }\label{table}
	\vspace{-6mm}
\end{table*}

In this section, we present (1) the main functionalities of \extra\ to explain three different team recommendation scenarios
and (2) key algorithms that support the main functionalities. 
\vspace{-3mm}
\subsection{System Functionality}\label{functionality_section}
The system provides intuitive visual explanations for the team recommendation results through the lens of the underlying network. Table~\ref{table} summarizes the main functionalities, where the rows represent explanations from three different perspectives (e.g., edges, nodes and attributes) for three team recommendation scenarios.

\noindent\textbf{A -- Explaining Team Replacement:} \extra\ identifies a few key (1) edges (the relationship between the candidate and other team members), (2) nodes (other existing team members) and (3) attributes (the skills of the candidate) that make the candidate and the departure member most similar. In this way, it could help the end-user (e.g., the team leader) understand why the underlying replacement algorithm thinks the given candidate is potentially a good replacement. \\
\textbf{B -- Explaining Team Expansion:} it provides the explanations from the following aspects, including (1) what are the unique new skills s/he brings to the team (i.e., attribute); and (2) what are the key collaborations the new team member might establish (i.e., edges) and with whom (i.e., nodes). \\
\textbf{C -- Explaining Team Shrinkage:} the prototype system flags the {\em absent} skills and connections with existing team members that makes the candidate most insignificant to the current team. In other words, we want to understand why the dismissal of this particular candidate would impose the least negative impact on the team.

\vspace{-3mm}
\subsection{Overview of Algorithm}\label{algorithms}

\noindent\textbf{A -- Graph Kernel for Team Recommendation: } random walk graph kernel provides a natural way to measure the similarity between two graphs 
~\cite{borgwardt2005protein} (e.g., the two team networks before and after a replacement) and turns out to be the core building block behind a variety of team recommendation scenarios. For example, in \emph{team replacement}, the objective is to find a \emph{similar} person $m$ to replace an existing team member $r$ in the context of the team itself. A good replacement $m$ should have a similar skill set as well as a similar collaboration structure with the existing team member. With graph kernel, we aim to find a candidate that makes the new team most similar to the old team.
For details regarding other scenarios (i.e., team expansion and shrinkage), please refer to \cite{borgwardt2007fast,li2017enhancing}.

\noindent\textbf{B -- Explaining Team Recommendation:} we seek to understand the team recommendation results via the influence of various graph elements to the corresponding graph kernels. We define the influence score of one specific element as the {\em rate} of the change in graph kernel $k(\mathbf{G}, \mathbf{G'})$. Given random walk graph kernel for labelled graph~\cite{li2015replacing}, the influence of a graph edge $W_{ij}$ can be computed as
\begin{equation}\label{influence}
	  \mathcal{I}\left( W_{ij} \right)  = \frac{\partial k\left( \mathbf{G},\mathbf{G'} \right)}{\partial W_{ij}} = c\mathbf{q_{\times}}^{T}\mathbf{R}\mathbf{L_{\times}}\left( (\mathbf{J}^{i,j} + \mathbf{J}^{j,i})\otimes\mathbf{W'}\right)\mathbf{R}\mathbf{L_{\times}}\mathbf{p} \nonumber
\end{equation}
\noindent where $\mathbf{R} = \left( \mathbf{I} - c~\mathbf{W_{\times}} \right)^{-1}$ \hide{, whose computation can be accelerated using power method,} and $\mathbf{J}^{i,j}$ is a single-entry matrix
with one at its $(i,j)$-th entry and zeros everywhere else.
In the scenario of team member replacement, an ideal candidate should have the key collaborations (with high influence score) as the departure member does, indicating their similarity in terms of how they collaborate with existing members.

The node influence of the $i^{\textrm{th}}$ member is defined as the aggregation of the influence of all the edges incident to this node, i.e., $\mathcal{I}\left(i\right) = \sum_{j | (i,j)\in \mathbf{E}}\mathcal{I}\left(W_{ij}\right)$. 
Existing team members with the highest node influence scores are expected to be key members in the team. 
\hide{Therefore a good candidate should also collaborate with these key members as the departure member does after replacement.}
\hide{In team expansion, these are the key team members the new member will work with. In team shrinkage, these are the key members that the candidate should have collaborated with.} 

Likewise, to compute the influence of a team member's attributes on the graph kernel, we take the derivative of the graph kernel w.r.t. the member's skill. Denote the $k^{\textrm{th}}$ skill of the $i^{\textrm{th}}$ team member in graph $\mathbf{G}$ as $L_{k}\left( i \right)$, the attribute influence can be calculated as
\begin{equation}\label{attribute}
  \mathcal{I}\left(L_{k}\left( i \right)\right) = \frac{\partial k\left( \mathbf{G},\mathbf{G'} \right)}{\partial L_{k}\left( i \right)} = \mathbf{q_{x}}^{T}\mathbf{R}\left( \frac{\partial \mathbf{L_{\times}}}{\partial L_{k}\left( i \right)}\right)\left( \mathbf{I} + c(\mathbf{W}\otimes \mathbf{W}')\mathbf{R}\mathbf{L_{\times}} \right)\mathbf{p_{\times}} \nonumber
\end{equation}
\noindent where $\frac{\partial \mathbf{L_{\times}}}{\partial L_{k}\left( i \right)} = \textrm{diag}\left( \mathbf{e}_i \right) \otimes \textrm{diag}\left(\mathbf{L'}\left(:, k\right)\right)$ and $\mathbf{e}_i$ is an  $n \times 1$ vector with one at the $i^{\textrm{th}}$ entry and zeros everywhere else.
Skills with the highest influence scores are the key skills that (1) the candidate and the departure member have in common in team replacement, (2) the new team member might bring in team expansion, and (3) the dismissal member might lack in team shrinkage.



	\vspace{-2mm}
\section{System Demonstration}

\begin{figure}[ht!]\label{fig:example}
  \centering
  \includegraphics[width=0.48\textwidth]{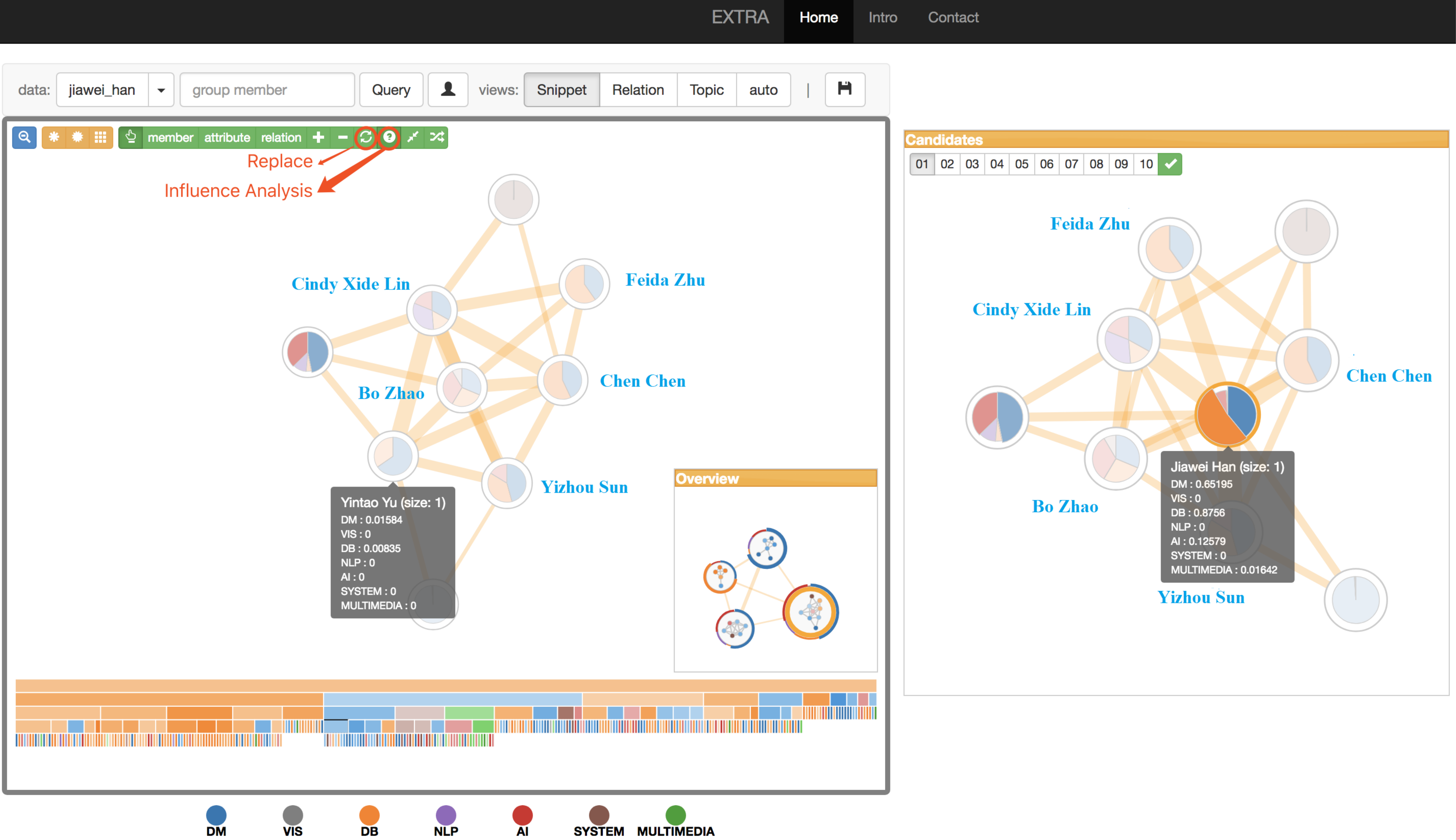}
  \vspace{-5mm}
  \caption{An illustrative example of influence analysis.}\label{experiment}
\end{figure}


\hide{\lli{can we mark the nodes w/ their last names in the figure instead of A,B,C..}}
Figure~\ref{experiment} presents the user interface of \extra\ and an example of visualization of influence analysis for team member replacement on a co-authorship network\footnote{A demo video of the system is available at \href{https://youtu.be/D4gcI-QHtps}{https://youtu.be/D4gcI-QHtps}. \\ System website: \href{http://144.202.123.224/system.html}{http://144.202.123.224/system.html }.}, which suggests that Dr. \emph{Jiawei Han} is the best replacement. The influence scores of all edges for both graphs are calculated by our proposed algorithm in Section~\ref{algorithms} and the width of edge is proportional to the influence score. The top-{\em 4} most influential edges are those that connect Dr. {\em Yu} with the 4 existing members, \emph{Lin, Zhao, Chen and Sun}, all of whom are considered as the key members and they also have strong collaborations with Dr. {\em Han}. After replacement, the top-{\em 5} most influential edges with Dr. {\em Han} overlap with those of Dr. {\em Yu} in the same ranking order. The only exception is that no edge exists between Dr. {\em Yu} and Dr. {\em Zhu} because there is no prior collaboration between them. In addition, the system can provide explanations from the attribute perspective. The key skills (represented in pie chart) shared by Dr. {\em Han} and Dr. {\em Yu} are {\it databases} and {\it data mining} in this case, which makes the team replacement recommendation more understandable. 

	\vspace{-3mm}
\section{Conclusions and future work}

In this demo, we present an interactive prototype system, \extra, as the first step towards explaining team recommendation through the lens of the underlying network where teams embed. \hide{The key algorithmic idea is to identify the most influential network substructures and/or attributes that account for the team recommendation results.} The system can provide intuitive visual explanations from different perspectives for various team recommendation scenarios.
Future work includes applying the proposed algorithm to other networked recommendation scenarios, e.g., user-product networks to explain why a certain product is recommended to a given user.



	\vspace{-2mm}
	\begin{acks}
		This work is supported by NSF (IIS-1651203, IIS-1715385, CNS-1629888 and IIS-1743040), DTRA (HDTRA1-16-0017), ARO (W911NF-16-1-0168), DHS (2017-ST-061-QA0001) and NSFC (61602306).
	\end{acks}
	\vspace{-3mm}
	\bibliographystyle{ACM-Reference-Format}
	\bibliography{references}


\begin{thebibliography}{4}


\ifx \showCODEN    \undefined \def \showCODEN     #1{\unskip}     \fi
\ifx \showDOI      \undefined \def \showDOI       #1{#1}\fi
\ifx \showISBNx    \undefined \def \showISBNx     #1{\unskip}     \fi
\ifx \showISBNxiii \undefined \def \showISBNxiii  #1{\unskip}     \fi
\ifx \showISSN     \undefined \def \showISSN      #1{\unskip}     \fi
\ifx \showLCCN     \undefined \def \showLCCN      #1{\unskip}     \fi
\ifx \shownote     \undefined \def \shownote      #1{#1}          \fi
\ifx \showarticletitle \undefined \def \showarticletitle #1{#1}   \fi
\ifx \showURL      \undefined \def \showURL       {\relax}        \fi
\providecommand\bibfield[2]{#2}
\providecommand\bibinfo[2]{#2}
\providecommand\natexlab[1]{#1}
\providecommand\showeprint[2][]{arXiv:#2}

\bibitem[\protect\citeauthoryear{Borgwardt, Ong, Sch{\"o}nauer, Vishwanathan,
  Smola, and Kriegel}{Borgwardt et~al\mbox{.}}{2005}]%
        {borgwardt2005protein}
\bibfield{author}{\bibinfo{person}{Karsten~M Borgwardt},
  \bibinfo{person}{Cheng~Soon Ong}, \bibinfo{person}{Stefan Sch{\"o}nauer},
  \bibinfo{person}{SVN Vishwanathan}, \bibinfo{person}{Alex~J Smola}, {and}
  \bibinfo{person}{Hans-Peter Kriegel}.} \bibinfo{year}{2005}\natexlab{}.
\newblock \showarticletitle{Protein function prediction via graph kernels}.
\newblock \bibinfo{journal}{{\em Bioinformatics\/}} \bibinfo{volume}{21},
  \bibinfo{number}{suppl\_1} (\bibinfo{year}{2005}), \bibinfo{pages}{i47--i56}.
\newblock


\bibitem[\protect\citeauthoryear{Borgwardt, Schraudolph, and
  Vishwanathan}{Borgwardt et~al\mbox{.}}{2007}]%
        {borgwardt2007fast}
\bibfield{author}{\bibinfo{person}{Karsten~M Borgwardt},
  \bibinfo{person}{Nicol~N Schraudolph}, {and} \bibinfo{person}{SVN
  Vishwanathan}.} \bibinfo{year}{2007}\natexlab{}.
\newblock \showarticletitle{Fast computation of graph kernels}. In
  \bibinfo{booktitle}{{\em NIPS}}. \bibinfo{pages}{1449--1456}.
\newblock


\bibitem[\protect\citeauthoryear{Li, Tong, Cao, Ehrlich, Lin, and Buchler}{Li
  et~al\mbox{.}}{2015}]%
        {li2015replacing}
\bibfield{author}{\bibinfo{person}{Liangyue Li}, \bibinfo{person}{Hanghang
  Tong}, \bibinfo{person}{Nan Cao}, \bibinfo{person}{Kate Ehrlich},
  \bibinfo{person}{Yu-Ru Lin}, {and} \bibinfo{person}{Norbou Buchler}.}
  \bibinfo{year}{2015}\natexlab{}.
\newblock \showarticletitle{Replacing the irreplaceable: Fast algorithms for
  team member recommendation}. In \bibinfo{booktitle}{{\em WWW}}. International
  World Wide Web Conferences Steering Committee, \bibinfo{pages}{636--646}.
\newblock


\bibitem[\protect\citeauthoryear{Li, Tong, Cao, Ehrlich, Lin, and Buchler}{Li
  et~al\mbox{.}}{2017}]%
        {li2017enhancing}
\bibfield{author}{\bibinfo{person}{Liangyue Li}, \bibinfo{person}{Hanghang
  Tong}, \bibinfo{person}{Nan Cao}, \bibinfo{person}{Kate Ehrlich},
  \bibinfo{person}{Yu-Ru Lin}, {and} \bibinfo{person}{Norbou Buchler}.}
  \bibinfo{year}{2017}\natexlab{}.
\newblock \showarticletitle{Enhancing team composition in professional
  networks: Problem definitions and fast solutions}.
\newblock \bibinfo{journal}{{\em TKDE\/}} \bibinfo{volume}{29},
  \bibinfo{number}{3} (\bibinfo{year}{2017}), \bibinfo{pages}{613--626}.
\newblock


\end{thebibliography}
	
\end{document}